# Selenium / Tellurium Two-Dimensional Structures: from Isovalent Se Dopants in Te to Atomically Thin Se Films

Guangyao Miao[1], Nuoyu Su[1,2], Ze Yu[1,2], Bo Li[1], Xiaochun Huang[1], Weiliang Zhong[1,2], Qinlin Guo[1], Miao Liu[1], Weihua Wang*[1], and Jiandong Guo*[1,2]

[1] Beijing National Laboratory for Condensed Matter Physics and Institute of Physics, Chinese Academy of Sciences, Beijing 100190, China
[2] School of Physical Sciences, University of Chinese Academy of Sciences, Beijing 100049, China

* Emails: weihuawang@iphy.ac.cn; jdguo@iphy.ac.cn.

**Abstract** Two-dimensional (2D) elemental semiconductors have great potential for device applications, but their performance is limited by the lack of efficient doping methods. Here, combining molecular beam epitaxy, scanning tunneling microscopy/spectroscopy, X-ray photoelectron spectroscopy, and density functional theory calculations, we investigate the evolution of the structural and electronic properties of 2D selenium/tellurium films with increased Se dosages on graphene/6H-SiC(0001) substrates. We found that Se atoms form isovalent dopants by replacing surface Te atoms, which introduces efficient electron doping and lowers the work function of Te films. With the Se dosage increasing, two types of elemental 2D crystalline Se structures, trigonal Se and $Se_8$ molecular assembly films, are obtained on ultrathin Te films, which are distinct from the amorphous Se acquired by depositing Se directly on graphene/6H-SiC(0001). Our results shed light on tuning the electronic properties of 2D elemental semiconductors by isovalent doping and constructing heterostructures of isovalent 2D elemental materials.

# 1. Introduction

The semiconducting two-dimensional (2D) materials such as transition metal chalcogenides and elemental atomic crystals have shown great potential for next-generation electronic and optical devices [1-5]. Absence of dangling bonds and their structural stability would help to overcome the short channel effect which hinders the miniaturization of transistors based on 2D materials, and lengthen the active time of Moore's law in semiconductor industry. For the low-dimensional semiconductors, doping is a key solution to tune their work function and lower the Schottky barrier between the metallic electrode and semiconductors, showing important value in device applications [6-9]. However, the traditional doping techniques always introduce non-negligible disorders in charge/lattice and Fermi-level pinning effect, thus degrade the devices' performance [7, 10-11]. So far, doping and band engineering of ultrathin van der Waals materials have been achieved by coating/intercalating charge transferred molecules/atoms [12-16], thickness engineering [17-18], surface functionalization [19-21], and fabricating atomic substitutional dopants [22-25]. Introducing isovalent substitutional dopants with same group elements brings much less distortions to the lattice, and has been widely used in fundamental studies and device developments based on low-dimensional materials [23, 26-28]. In low-dimensional materials, the atomic structure and electronic properties of isovalent dopants are directly accessible by both local probing and surface sensitive techniques, providing us opportunities to investigate the doping mechanisms of isovalent dopants both at the atomic and macroscopic scales.

Trigonal tellurium is one of the elemental semiconducting materials with many excellent physical properties such as outstanding thermoelectricity [29-31], high carrier mobility [3, 32-34], mid-infrared polarized photo response [35-37], spin-polarized band structure and Weyl fermions [38-41], and novel electronic-magnetic coupling effect [42-43], which have attracted enormous research interest in the past few years. The pure Te crystal shows a *p*-type semiconducting character [38, 44-45]. Therefore, an efficient electron doping method was in demand for Te-based fundamental studies and device exploitations. Up to now, the electron doping to bulk Te crystals and thick Te films has been realized by potassium deposition and atomic layer deposition dielectric doping technique [39, 46]. When the thickness of tellurium approaches the ultrathin/monolayer limit, the tellurium shows multiple phase character [47], thickness-dependent bandgap (0.33 eV~1.0 eV) [32, 48], tunable carrier type [44], and low-dimensional edge states [49]. These properties make ultrathin tellurium films prospective applicant in future energy and information devices. However, the bulk doping methods are hard to realize efficient doping while keeping the lattice intact in ultrathin Te films. Huang *et al.* reported the electron doping from the few-layer graphene/6H-SiC(0001) (graphene/SiC) substrate to the monolayer Te films [44]. To cast off the constrain of the substrate, efficient doping methods and understanding the doping mechanisms are indispensable for both fundamental research and device engineering of ultrathin Te films in a wider context.

Here, combining molecular beam epitaxy (MBE), scanning tunneling microscopy and spectroscopy (STM/STS), X-ray photoelectron spectroscopy (XPS), and density functional theory (DFT) calculations, we investigate the evolution of structural and electronic properties of ultrathin Te films with increasing Se dosage. Upon Se deposition on ultrathin Te films and post-annealing, Se atoms substitute for surface Te atoms, forming $Se_{Te}$ substitutional dopants in the film while maintaining the helical structure. As revealed by our STM, STS and XPS experiments, the as-prepared *p*-type semiconducting Te film is progressively tuned to *n*-type by increasing the concentration of $Se_{Te}$ dopants, indicating electron doping effect of $Se_{Te}$ dopants in ultrathin Te films. This electron doping effect is corroborated with the charge density redistribution as calculated by our first-principles calculations. By further increasing the Se dosage, we further prepare ultrathin trigonal Se films and $Se_8$ molecular films on tellurium films, forming vertical heterostructures of 2D Se/Te. Our work provides insight into the doping effect of isovalent dopants, and sheds light on the high-precision doping of 2D semiconducting elemental materials and fabrications of novel elemental 2D crystals.

## 2. Results and discussions

### 2.1 Evolution of Se/Te 2D structures

Trigonal tellurium and selenium crystalline films are isostructural, i.e., both are composed of parallelly arranged 3-fold helical chains, with right-handed for space group $P3_121$ and left-handed for space group $P3_221$[41]. Here, we firstly grow ultrathin trigonal Te films on graphene/SiC substrate following the reported methods [48-49]. Then Se atoms were introduced to the surface of Te films by controlled Se deposition. Figure 1**a** shows the sketch of the fabrication of Se/Te structures in our experiments. Upon deposition of Se on ultrathin Te films held at room temperature, the Se atoms tend to form amorphous clusters near the edges of Te films, as shown in the left panel of Fig.1**a** and Fig.1**b**. After post-annealing at ~60 °C in ultrahigh vacuum, the Se clusters disappeared, and the Se atoms substitute for surface Te atoms, forming $Se_{Te}$ substitutional dopants. As the structural sketch in the middle panel of Fig.1**a** and the STM image in Fig.1**c** show, the bright and dim atoms (Te and Se atoms, respectively) are distributed randomly on the surface layer and the film maintains the helical structure. If we further increase the Se dosage, two types of 2D selenium crystalline structure (right panel of Fig.1**a**) are fabricated on the tellurium films. Their STM images are displayed in Figs.1**d** and 1**e**.

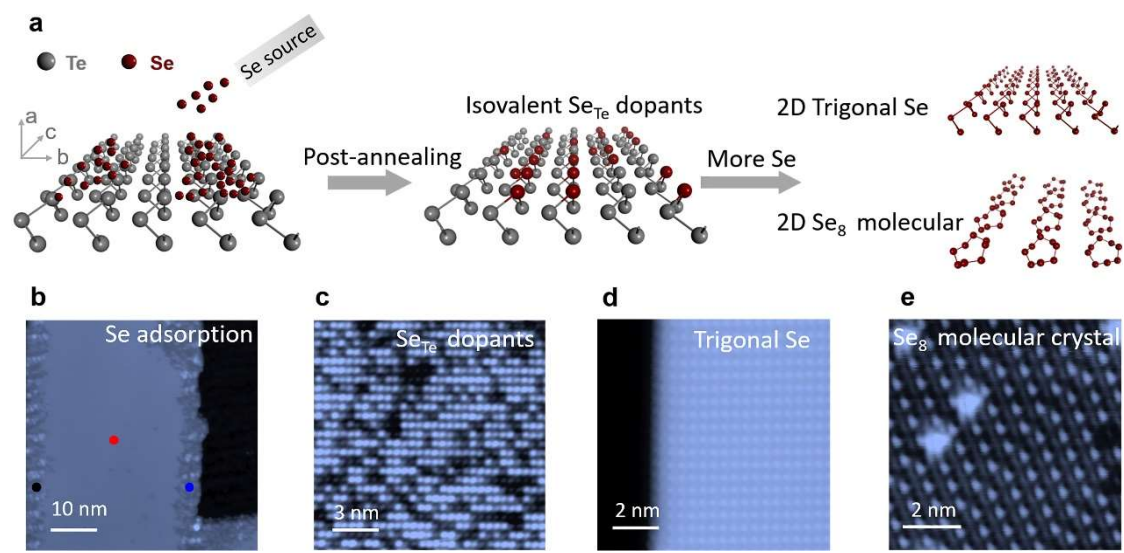

Figure 1 Fabrication of Se/Te 2D structures. **a** Sketch of the evolution of the Se/Te 2D structures. Te and Se atoms are represented by gray and wine balls, respectively. **b**, **c**, **d** and **e** STM images of the Te film after depositing a small amount of Se, Te films with surface $Se_{Te}$ dopants after post-annealing, ultrathin trigonal Se films, and $Se_8$ molecular crystal films. Scanning parameters: **b**: $V_b$= -2.0 V, $I_t$= 50 pA; **c** $V_b$= -2 V, $I_t$= 100 pA; **d**: $V_b$= -2 V, $I_t$= 50 pA; **e**: $V_b$= -2 V, $I_t$= 50 pA.

### 2.2 Isovalent $Se_{Te}$ dopants

Figures 2**a** to 2**d** compare the atom-resolved STM images of pure Te and Se-doped Te films with increased Se dosage after post-annealing. Different from the homogenous

contrast on pure Te film (Fig. 2**a**), the STM images acquired on the Te films with surface Se$_{Te}$ dopants show randomly distributed bright and dim atoms (Figs. 2**b-d**). Since the ratio of dim to bright atoms in the Te film increases with Se dosage, the dim atoms are assigned to Se$_{Te}$ dopants. The different contrast of Te and Se atoms are given by their different covalent radii. To confirm the assignment, we further simulated the STM image of trigonal Te film with Se$_{Te}$ dopants by DFT calculation. The simulated STM image is shown in the inset of Fig. 1**b**, in which the Se atoms show a lower apparent height than Te atoms, evidencing that the dim atoms are the Se$_{Te}$ dopants. It should be noted that with the Se dosage increasing, the Se atoms may substitute for the Te atoms not only in the top-most sublayer, but also in the deeper sublayers.

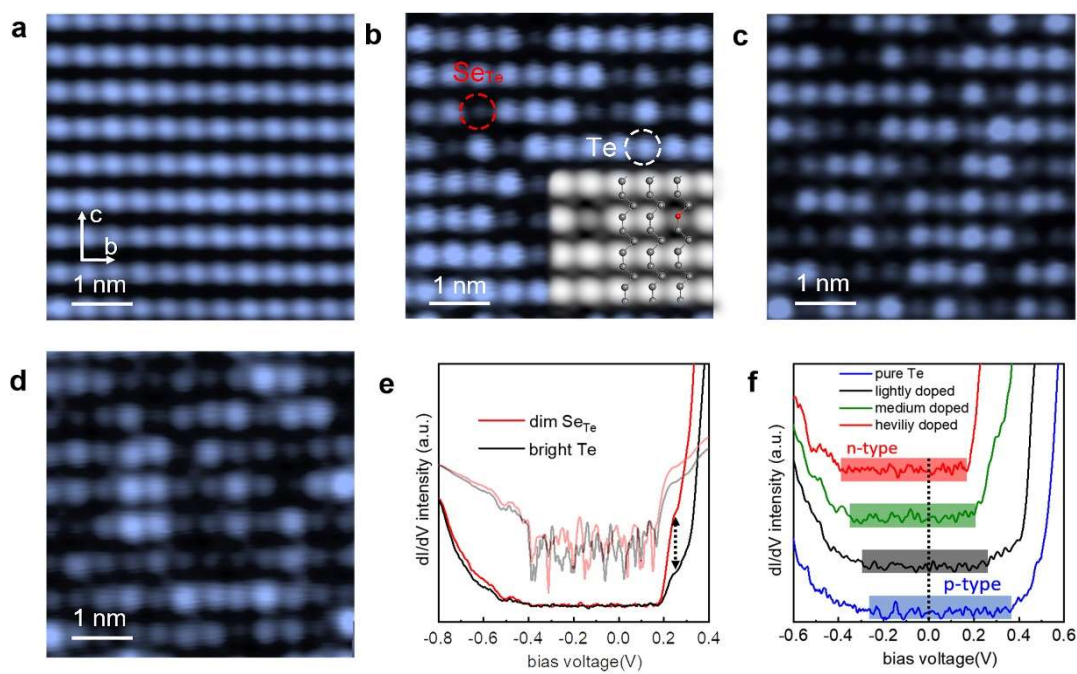

Figure 2 STM and STS Characterization of Te films with increased Se$_{Te}$ dopants. **a-d** Atom-resolved STM images of pure Te film (**a**), lightly Se-doped Te film (**b**), medium Se-doped Te film (**c**), and heavily Se-doped Te film (**d**). Scanning parameters: **a**, $V_b$= -1 V, $I_t$= 100 pA; **b**, $V_b$= -1 V, $I_t$= 100 pA; **c**, $V_b$= -1 V, $I_t$= 100 pA; **d**, $V_b$= -1 V, $I_t$= 100 pA. The DFT simulated STM image of Se$_{Te}$ isovalent dopant at $V_b$= -1 V is superposed on (**b**) together with a structural model. The red and grey ball represent the Se and Te atoms, respectively. **e** Site-dependent differential conductance (d$I$/d$V$) spectra measured on the bright atoms (Te) and dim atoms (Se$_{Te}$ substitutional dopant), the translucent curves show their logarithmic spectra. **f** d$I$/d$V$ spectra measured on the pure Te film (blue curve), lightly doped film (surface Se/Te ratio at ~1:4, black curve), medium doped film (surface Se/Te ratio at ~1:1, olive curve), and heavily doped film (surface Se/Te ratio at ~3:1, red curve). The curves are vertically shifted for clarity. The energy gaps of the spectra are highlighted by different colors.

By measuring site-dependent STS, we investigate the effect of Se doping on the electronic states of Te films. The typical d$I$/d$V$ spectra measured on Se$_{Te}$ and Te atoms are shown in Fig. 2**e**. The spectrum acquired on the Se$_{Te}$ site shows enhanced local density of states (LDOS) near the conduction band minimum (CBM) compared to the

Te site, while the energy positions of valence band maximum (VBM) and CBM are the same. No in-gap state is observed on Se$_{Te}$ dopants, similar to the O$_{Se}$ and O$_S$ substitutional dopants in MoSe$_2$ and WS$_2$ [50]. However, the Se$_{Te}$ substitutional dopants show different behavior compared with O$_{Se}$ and O$_S$ that introduce resonance states below the VBM [50]. Our DFT calculations reproduced the enhanced state near the CBM on Se$_{Te}$ site, as shown in Fig.S1(Supplementary Fig. 1). Moreover, both the Se$_{Te}$ site and Te site display *n*-type semiconducting behavior in contrast to the *p*-type semiconducting character of pure Te films. In Fig. 2**f**, we compare the d*I*/d*V* spectra measured on pure Te films and Se-doped Te films with increased Se dosage. One can see that the *p*-type pure Te is progressively tuned to *n*-type with increased Se dosage, evidencing the electron doping effect of Se$_{Te}$ dopants. (Note: To avoid the doping effect of substrate, all the STS data are acquired on Te films with the thickness of ~12 layers.) The doping effect of the Se$_{Te}$ dopants has a large penetration depth. As shown in Fig.S2 (Supplementary Fig. 2), the area with large amounts of Se$_{Te}$ dopants and nearly without Se$_{Te}$ dopants are both *n*-type semiconducting, which implies that the electron doping effect of the Se$_{Te}$ dopants is delocalized and free of Fermi-level pinning effect.

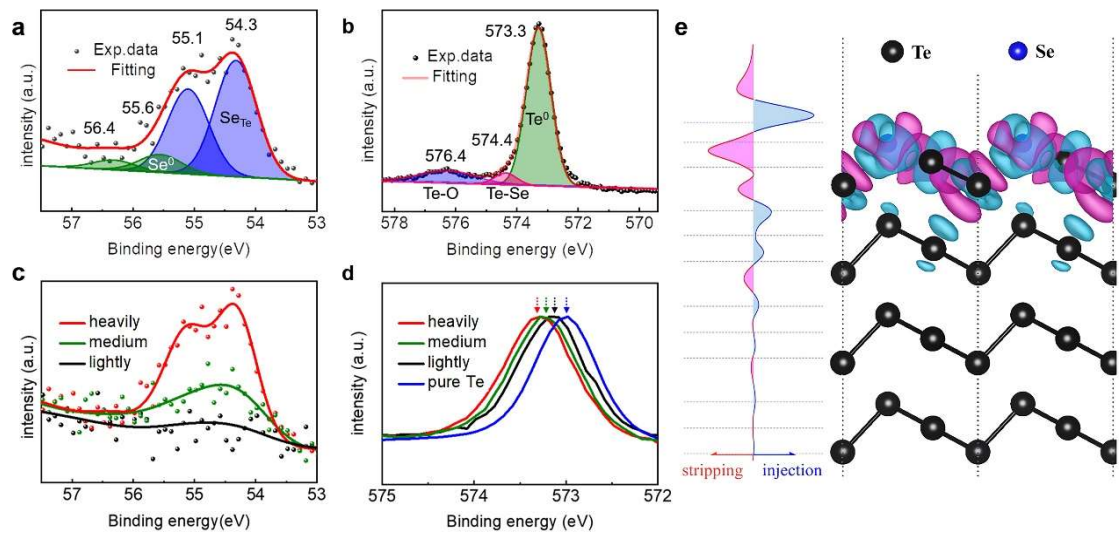

Figure 3 XPS characterization and DFT calculation. **a**, **b** Survey near the 3$d_{5/2}$ and 3$d_{3/2}$ peak of Se (**a**) and Te (**b**) in the heavily doped sample. **c** Comparation of the XPS results of three doped samples with different Se dosage. **d** Evolution of the Te-3$d_{5/2}$ peak with increased Se dosage. The centroids of the peaks are indicated by colored arrows following the color code in (**c**). **e** DFT calculated charge redistribution at the surface of the Te film when Se$_{Te}$ substitutional dopants are introduced, the left panel shows the electron stripping and injection situation along the film normal direction.

To characterize the doping effect of the isovalent Se$_{Te}$ dopants at the macroscale, we carry out XPS survey on pure Te films and Se-doped Te films. As shown in Fig. 3**a**, the binding energy (BE) of Se 3$d_{5/2}$ and 3$d_{3/2}$ in the heavily doped sample are shifted to 55.1 eV and 54.3 eV, ~1.3 eV lower than that in pure Se (55.6 eV and 56.4 eV, respectively), indicating the negative chemical states of Se$_{Te}$ dopants. By fitting the Te 3$d_{5/2}$ spectrum of heavily doped sample in Fig. 3**b**, three components are resolved: the weak feature at 576.4 eV (blue area) is assigned to oxidized Te$^{2+}$; the component at 574.4 eV, which is ~1.4 eV higher than that in pure Te, is assigned to those Te atoms bonded with Se atoms; the main peak at 573.3 eV is assigned to $Te^0$ atoms. Compared to pure Te, the BE of $Te^0$ atoms in the heavily doped sample is 0.3 eV higher, which is attributed to a decreased work function caused by the electron doping effect of $Se_{Te}$ dopants. Consequently, the BE of the main peak given by $Te^0$ atoms can be substantially affected by the Se dosage. The Se $3d$ and Te $3d_{5/2}$ spectra in doped Te films with different Se dosages are compared in Figs. 3**c** and 3**d**, respectively. (In Fig. 3**d** the Te $3d_{5/2}$ spectrum in pure Te is also plotted for comparison.) With increased Se dosage, the Se $3d$ signal get stronger (Fig. 3**c**), and the main peak center of Te $3d_{5/2}$ signal shifts from 573.0 eV in pure Te, to 573.3 eV in heavily Se doped Te (Fig. 3**d**), which is consistent with the STS results (Fig. 2**f**). To exclude the energy shift originates from the measurement error, we compare the XPS survey of the C-1s in these samples and no energy shift is observed (Supplementary Fig. 3). Our XPS experiments indicate that the BE of the Te $3d_{5/2}$ main peak may serve as an indicator for the electron doping level in doped Te films.

To figure out the micro-scale physical origin of the doping effect caused by the isovalent $Se_{Te}$ dopants in Te films, we performed first-principles calculations to investigate the change in charge density before and after substituting Se for Te. The differential charge density (DCD) of the system is calculated by

$$\Delta\rho_{Se_xTe_{1-x}} = \rho_{Se_xTe_{1-x}} - \rho_{Se} - \rho_{Te}$$

where $\rho_{Se_xTe_{1-x}}$, $\rho_{Se}$ and $\rho_{Te}$ represent the charge densities of the Se-doped Te, Se atom and Te films, respectively. Additionally, we also calculated the one-dimensional planar averaged DCD along the film normal direction by integrating the DCD along the film in-plane direction. Figure 3e displays the calculation results, revealing that $Se_{Te}$ dopants lead to charge density redistribution, introducing ripples in DCD penetrating from the Se towards the inner layers of the film, at least two layers deep. Electron accumulation occurs at the Se atoms, causing electron depletion at the Te atoms directly bonding to the Se atoms. This observation is consistent with the 1.3 eV BE shift seen in XPS experiments. To balance the positive charge at the Te atoms bonding to Se, electrons are also accumulated in the undoped second layer. The electron doping effect of Se dopants in Te films can be rationalized as follows: Upon the occurring of the Se substitution, the Se atoms are negatively charged, while the Te atoms directly bonding to Se atoms are positively charged due to their different electronegativity. In contrast, the Te atoms near the doped region are negatively charged due to screening effect. This redistribution of charge density near the surface induced by $Se_{Te}$ dopants lead to the lowering of the Te films' work function. Which could explain that the increasing Se dosage in Te films would drive the shift of $Te^0$ $3d_{5/2}$ and $3d_{3/2}$ BE towards higher energy observed in the XPS experiments (Fig. 3**d**), and transition from *p*-type to *n*-type semiconductors observed in the STS measurements (Fig. 2**f**). This mechanism to tune the work function could also be applied to other elemental 2D semiconducting materials.

**2.3 2D Selenium structures**

By further increasing the Se dosage, we successively grow ultrathin thin Se films on Te films. Two types of crystalline Se films are observed in STM images, the trigonal Se films and molecular films. The trigonal Se film has similar structure as trigonal Te, but with different lattice parameters, as shown in Fig.4**a**. Figures 4**b** and 4**c** are the large-scale and zoomed-in STM images of trigonal Se, respectively, showing in-plane lattice constants $b$=0.45 nm and $c$= 0.55 nm, and a nominal height of 0.41 nm for trigonal Se layer. Figure 4 **e** shows the large-scale STM image of Se molecular film, indicating a nominal height of 0.45 nm. As shown in Fig. 4**f**, the molecular film shows a unit cell of $a$=0.81nm, $c$=1.05 nm and an angle of 111º between them. A tentative structural model of the molecular film, calculated by the material database Atomly,[51] is shown in Fig. 4**d**. The molecular film is composed of $Se_8$ molecules with in-plane lattice constants of $a_0$= 0.89 nm, $c_0$= 1.12 nm and an angle of 105º between them, in good agreement with our STM results. As compared in Fig. 4**g**, the bandgap of trigonal Se is 1.68 eV, and the molecular film shows a bandgap of 1.47 eV, both exceeding the bandgap in Te film and in consistent with the calculation results.[51] Considering the trigonal Se film and molecular film are both weak *n*-type semiconductors according to their d$I$/d$V$ spectra, they may form vertical heterojunctions junctions with the underlying Te films.

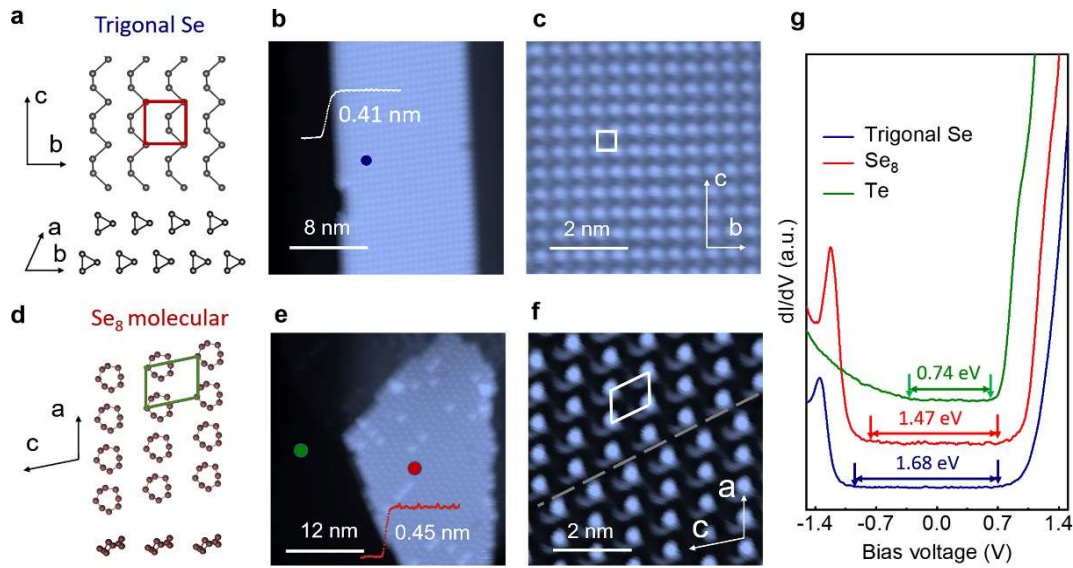

Figure 4 2D elemental selenium structures. **a** Crystal structure of trigonal Se. The upper panel and lower panel show the top view and side view respectively. **b** and **c** Large-scale and atom-resolved STM images acquired on the ultrathin trigonal Se films. **d** Crystal structure of monolayer $Se_8$ films grown on Te films. The upper panel and lower panel show the top view and side view respectively. **e** and **f** Large-scale and atom-resolved STM image of monolayer $Se_8$ films. **g** Differential conductance spectra acquired on trigonal Se, monolayer $Se_8$, and Te films.

## 3. Conclusion

In summary, we achieved the controllable fabrication of surface $Se_{Te}$ dopants on the ultrathin Te films. The surface isovalent dopants introduce efficient electron doping to the ultrathin Te films and tune the ultrathin Te films from *p*-type to *n*-type without breaking the continuity of lattice and introducing Fermi-level pinning effect. The doping method has low processing temperature, efficient doping capacity, and is independent of the substrate, which could be compatible with the mainstream devices' preparation technology. We expect this method, combined with the phase mask technology, can be utilized in constructing in-plane *p-n* junctions based on ultrathin tellurium films without the constrain of the substrate. This doping method shows enormous applying potential in tellurium-based optical and electrical devices and may be extended to the vast 2D elemental semiconducting materials. The successfully growth of the ultrathin trigonal Se films and monolayer $Se_8$ films on Te films also paves the way to investigation of devices and physics at the ultrathin Se/Te interface.

## 4. Experimental Section

**Sample fabrications**: The experiments were carried out in an ultra-high vacuum molecular beam epitaxial (MBE) and scanning tunneling microscope (STM) system (Unisoku) with a base pressure lower than $2 \times 10^{-10}$ mbar. The SiC substrates were purchased from Tankeblue Semiconductor, Beijing. The 6H-SiC(0001) substrate was degassed at 660℃ in the MBE chamber for more than 12 hours. Then we annealed the SiC at about 1250℃ for 30 mins to get the bilayer graphene/SiC heterostructure with $6\sqrt{3} \times 6\sqrt{3}$ $R30°$ corrugations. The deposition of Te and Se was achieved by evaporating high-quality Te and Se powder (99.999%) by the Knudsen cell. The Te films were grown on Gr/SiC substrate following the reported method[1]

**Sample characterizations**: The as-grown samples were investigated by the in-situ STM at 78 K. The STM topographic images were acquired in constant-current mode with the bias voltage applied to the sample (Scanning at positive bias voltage detect the unoccupied states). To measure STS, the d$I$/d$V$ signals were acquired using a lock-in amplifier with a sinusoidal modulation of 1517 Hz at 10 mV. We typically acquired 10–20 spectra for the same site to verify for reproducibility and then averaged them to increase the signal-to-noise ratio. To confirm the chemical states of the samples, the samples were transferred to the chamber of the X-ray photoelectron spectroscopy (XPS) systems in 10min. The XPS experiments were conducted on the Thermo Scientific ESCALAB 250 at room temperature. The calibration of the binding energy (BE) of the XPS is calibrated with respect to the pure bulk Au $4f_{7/2}$(BE=84 eV) and Cu $2p_{3/2}$ (BE=932.7 eV) lines. The BE is referenced to the Fermi level ($E_f$) calibrated by using pure bulk Ni as $E_f$ = 0 eV.

**DFT calculations**: First-principles calculations based on density functional theory (DFT) implemented in the Vienna Ab initio Simulation Package (VASP)[2] were used throughout the work to model the material system. The projector augmented-wave (PAW) method[3] was used to describe the ion-electron interactions and the generalized gradient approximation (GGA)[4] within the Perdew-Burke-Ernzerhof (PBE) framework[5] was used as the exchange-correlation functional. All calculations use a plane wave cutoff of 520 eV and a resolution value of 0.03 between k-points in reciprocal space in units of 2π·Å$^{-1}$ to obtain converged results. The surface calculation of Te-Se is modeled by a slab supercell with a vacuum thickness of 20 Å. The thickness of Te layers in the slab supercell exceeds 25 Å to prevent the interaction between the two terminal surfaces. To simulate the case of Se doping, a Te at the top of the slab supercell was replaced by Se. The lattice parameters of the supercell are fixed at the bulk value in the in-plane direction, and all atoms are free to relax until the force is < 0.02 eV/Å. Spin polarization is not considered because the volume magnetic susceptibility of Te and Se is less than $1 \times 10^{-4}$. To obtain the differential charge density, we separately fixed the Te layers and the isolated Se atom at the same atomic position in the Te-Se slab supercell and calculated the charge density of the Te layers and the isolated Se atom, respectively. Furthermore, whether or not van der Waals interactions are considered gives the same physical picture. The simulated STM images are obtained by fixing the lattice parameters of the Te films at the experimental results.

# Acknowledgements

This work was supported by the National Natural Science Foundation of China (Grants No. 11974399, 11974402), the Strategic Priority Research Program of Chinese Academy of Sciences (Grants No. XDB33000000), G.M. acknowleges the support from the IOP-Humboldt postdoc fellowship in physics of institute of physics, CAS and Integrative Research Institute for the Sciences of Humboldt-Universität zu Berlin.

# Conflict of interest

The authors declare no conflict of interests.

# Data Availability Statement

The data that support the findings of this study are available from the corresponding author upon reasonable request.

## Supporting Information:

**DFT calculated DOS of Se$_{Te}$ dopant**

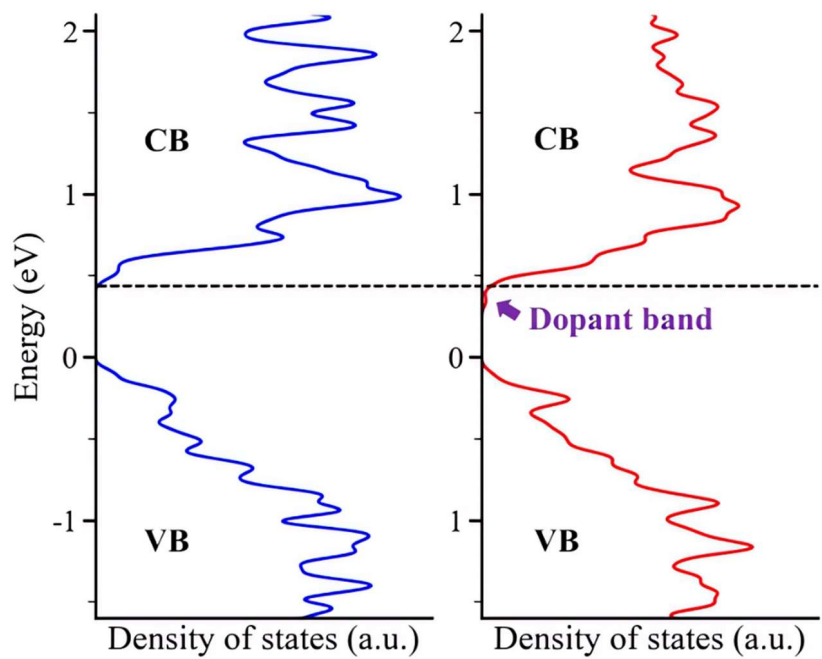

Figure S1 The DFT calculated density of states acquired at the pure Te and the Se$_{Te}$ dopant sites. The dopant band near the CBM are indicated by the purple arrow.

## More experimental results

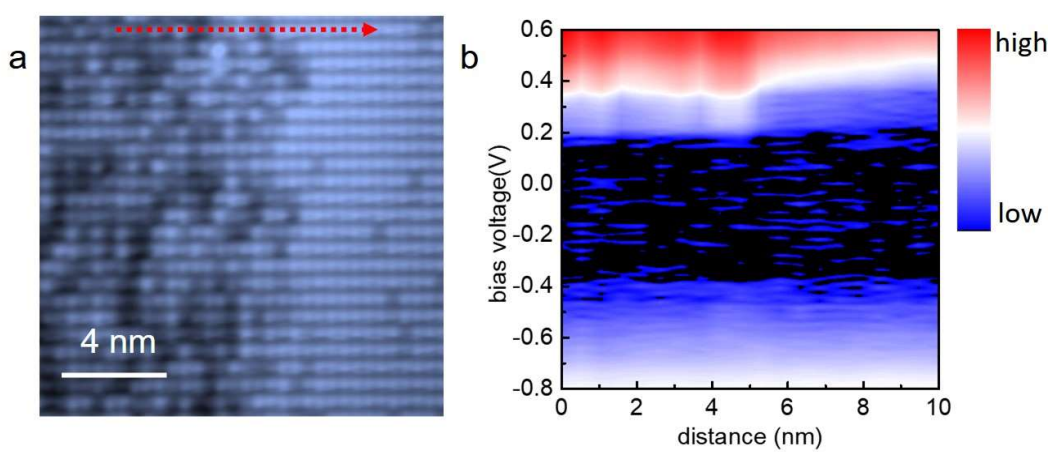

Figure S2 **a** The STM image ($V_b$= -1.0 V, $I_t$= 100 pA) acquired from a sample including the interface between the Se-doped and undoped areas. **b** Spatial resolved line dI/dV (logarithmic) mapping along the red dashed arrow in **a**.

Taking use of STS measurements, we investigate the evolution of electronic states with increased Se dosage, the Se atoms prefer to adsorb and substitute for Te atoms near the edge of Te films firstly and leave the inner region undoped, providing us an interface to reveal the electronic structure evolution from Se-doped area to undoped Te area by site-dependent STS. Figure S2**a** shows such an interface between Se-doped (left) and undoped (right) regions. We measured STS along the red dashed arrow (along the *b* axis of Te film) in Fig. S2**a**. As shown in Fig. S2**b**, in contrast to pure Te films an intrinsic *p*-type semiconductor, both the doped and undoped regions are *n*-type semiconducting. This indicates the Se$_{Te}$ dopants can introduce a delocalized electron doping effect on ultrathin Te films

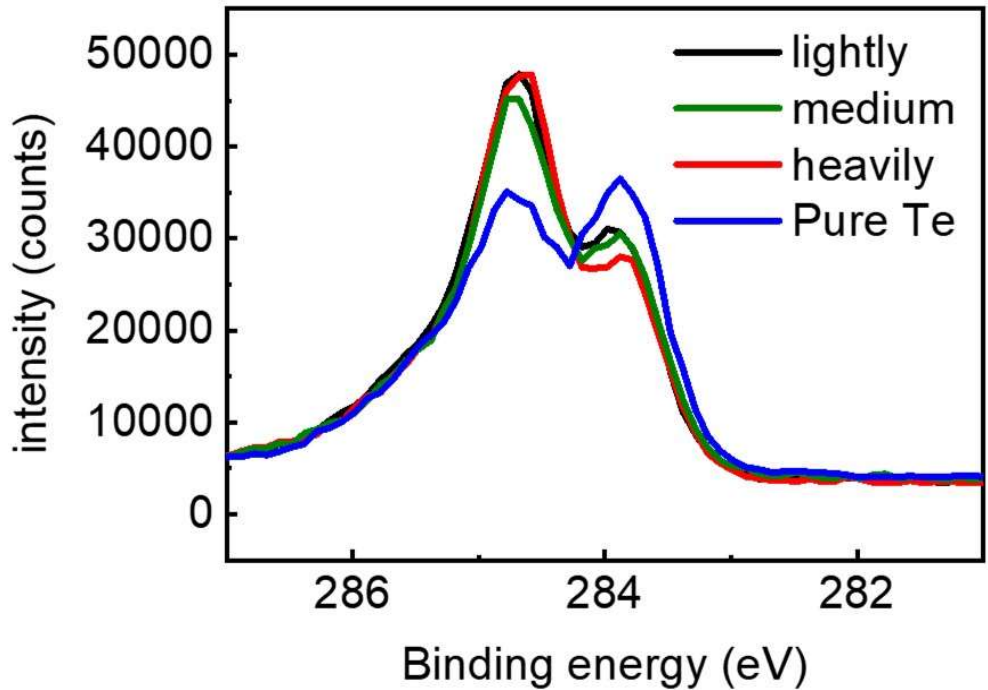

Figure S3 The X-ray photoemission spectroscopy survey of samples with different Se dosage near the binding energy of C-1s.

The C-1s peak centers of the samples with different Se dosage are nearly located at same energy position. The measured BE of C 1s are consistent with the reported results acquired on the few-layers graphene epitaxial growing on SiC [6]. These results validate the binding energy shift of Te-3$d_{5/2}$ signal with Se dosage in the main text.